\documentclass{article}
\usepackage{amsmath}
\usepackage{graphicx}
\usepackage{dcolumn}
\usepackage{bm}
\usepackage{epsfig}
\usepackage{graphics}
\usepackage{amsfonts}
\usepackage{amssymb}

\begin{document}

\title{Nonlinear absorption in discrete systems}
\author{A. Spire and J. Leon\\
Physique Math\'ematique et Th\'eorique, CNRS-UMR5825,\\
 Universit\'e Montpellier 2, 34095 Montpellier (France)}
\date{ }
\maketitle

\begin{abstract}  In the context of nonlinear scattering, a continuous wave
incident onto a  nonlinear discrete molecular chain of coupled oscillators can
be partially absorbed as a result of a 3-wave resonant interaction that couples
two HF-waves of frequencies close to the edge of the Brillouin zone. Hence both
nonlinearity and discreteness are necessary for generating this new absorption
process which manisfests itself by soliton generation in the medium. As a
paradigm of this {\em nonlinear absorption} we consider here the Davydov model 
that describes exciton-phonon coupling in hydrogen bonded molecular chains.
\end{abstract}

%\pacs{05.45.Yv}
%\submitto{\JPA}

\section{Introduction}

The scattering of waves becomes extremely rich when nonlinearity comes into
play, and one of its most fundamental effect is  {\em soliton  generation}
which results in energy localization and propagation.  For instance an incident
radiation on a two-level medium at the resonant  frequency can be totally
transmitted, instead of being absorbed, a property described  in \cite{sit} as
{\em self induced transparency}. It results from the nonlinear coupling of
radiation with medium population, a mechanism which generates the solitons,
vectors of energy transmission.

Other interesting  processes of nonlinear wave scattering are the two-photon 
propagation, second harmonic generation and stimulated Raman scattering.  In
that last case, the nonlinear interaction induces (laser) pump depletion  and
phase effects result in Raman spike generation (short duration pump  repletion)
\cite{druhl}. Although the Raman spike generation is not a solitonic effect,
here also the nonlinearity is the fundamental tool \cite{srs}.

Recently discovered, the {\em nonlinear supratransmission} \cite{alex,nst}  is
a nonlinear scattering process where a totally reflective medium switches  to
high transmissivity above some threshold  because of nonlinearity. 
The bifurcation takes its origin in a nonlinear instability \cite{instab}
which has a wide field of application in generic nonlinear evolutions
\cite{jphysc}.

Particularly interesting is the application of this concept to the scattering
of a continuous-wave laser beam incident onto a Bragg mirror which switches
from total reflection to transmission by means of gap soliton generation
\cite{yariv,expe}. It has been demonstrated that the switch is a manifestation
of nonlinear supratransmission which has  in particular allowed to compute an
analytic explicit expression of the  intensity flux threshold \cite{bragg}.

We are interested here in the concept of {\em nonlinear absorption} where a 
me\-dium, transparent to incident radiation in the linear regime, can become
absorbant under contribution of  nonlinearity which is due here  to coupling of
waves of different nature. Such is the case with the  Davydov model
\cite{davydov} describing the coupling of high frequency  optical phonon waves
to low frequency acoustic waves. A generic field of application of the Davydov
model is the cristallin acetanilide where the acoustic wave represents the
hydrogen bond between adjacent molecules, while the optical phonon wave
represents the C=O stretching (amide-I mode). This model has been widely
studied in its stationnary limit where the hydrogen bond is supposed to have a
time response much longer than the high frequency C=O stretching. In that
situation the resulting simplified model is the nonlinear Schr\"odinger
equation that has been used to interprete the anomalous IR-absorption band by
soliton  generation \cite{scott}.

The Davydov model has been recently reexamined in the context of resonant
two-wave interaction where a rigorous multiscale analysis transforms the
discrete model to a continuous integrable partial differential equation which
keeps the fundamental wave coupling effect and  possess multi-soliton solutions
\cite{flora}. This two-wave resonant interaction process takes into account an
incident high-frequency optical phonon wave that resonates with the
low-frequency acoustic wave as soon as the HF-wave group velocity equals the
LF-wave phase velocity. This is the Benney criterion \cite{benney}  that has
revealed its efficiency through numerical simulations of the Davydov model
\cite{kopidakis}. 

The two-wave resonant process assumes no backward propagation and thus neglects
the multiple reflections due to the periodicity of the cristal. We consider
here a three-wave interaction process  that couples the incident optical
phonon  wave both to the acoustic excitation and the reflected optical phonon
wave. We shall discover that an incident and a reflected high-frequency waves
can cooperate resonantly with a low-frequency acoustic wave thanks to the
discrete nature of the basic model.

We will demonstrate that this scattering process allows for absorption of
incident radiation by a purely nonlinear effect that generates a three-wave
soliton in the medium. The nonlinear effect is effectively an absorption as a
part of the incident energy flux is transfered to the medium.

\section{Three-wave interaction in the Davydov model}

\subsection{Basic model}
Our starting point is the Davydov model \cite{davydov} for the eigenstate
$a_n(t)$ of the amide-I excitation (the corresponding dynamical variable
represents the C=O stretching) coupled to the dynamical variable  $\beta_{n}$
that represents the longitudinal displacement along the hydrogen-bond chain.
It reads
\begin{align}
&  i\hbar\dot a_{n}=\left[ E_{0}+W+\chi(\beta_{n+1}-\beta_{n})\right]a_{n}
-J(a_{n+1}+a_{n-1}),\\
&  M\ddot\beta_{n}=K(\beta_{n+1}-2\beta_{n}+\beta_{n-1})+
\chi(|a_{n}|^{2}-|a_{n-1}|^{2}),
\end{align}
where $M$ is the mass of the peptide group, $K$ the
spring constant of the hydrogen bond, $E_{0}$ the energy of the C=O stretching
and $W$ the the total energy of the peptide group displacements. The constant
$\chi$ is the coupling parameter and $J$ measures the energy of the
dipole-dipole interaction of C=O stretching oscillations. Overdot stands for
derivation with respect to the physical time $T$.

Upon defining the dimensionless time $t=JT/\hbar$,
and the adimensional quantities
\begin{equation}\label{Q-B}
\Psi_{n}=a_{n}\frac{\hbar\chi}{J\sqrt{JM}}\ \exp[\frac{i}{J}(E_{0}
+W-2J)\ t\ ]\ ,\quad Q_{n}=\frac{\chi}{J}(\beta_{n+1}-\beta_{n})\ ,
\end{equation}
the system becomes
\begin{align}
&  i\partial_{t}\Psi_{n}+(\Psi_{n+1}-2\Psi_{n}+\Psi_{n-1})=
Q_{n}\Psi_{n}\ ,\label{dav-sco-psi}\\
&  \partial_{t}^{2}Q_{n}-V^{2}(Q_{n+1}-2Q_{n}+Q_{n-1})=|\Psi_{n+1}|^{2}
-2|\Psi_{n}|^{2}+|\Psi_{n-1}|^{2},\label{dav-sco-q}
\end{align}
which constitutes our basic example of a discrete nonlinear coupled wave
system of equations. Note that we are left with a single constant, the 
adimensional sound velocity
\begin{equation}
V=\frac{\hbar}{J}\ v_{p}\ ,\quad v_{p}=\sqrt{\frac{K}{M}}, \label{V-vp}
\end{equation}
where  $v_{p}$ the phonon velocity (cells per second), and that
the coupling parameter $\chi$  has been absorbed in the amplitude of 
$\psi_n(t)$.

\subsection{Multi-scale expansion}

Following the multiscale expansion method \cite{With} we assume a
representation of the solution $\{\Psi_{n}(t),\ Q_{n}(t)\}$ as formal series
in terms of a small parameter $\epsilon$ where space-time dependences occur
at a sequence of slow scales. Since it is not necessary to push the series at
arbitrary order, we write explicitely the first relevant terms only as
\begin{align}
&\Psi_{n}(t)=\epsilon \varphi^{(0)}(n_0,x_1,\cdots;t_0,t_1,\cdots)+
\epsilon^2\varphi^{(1)}(n_0,x_1,\cdots;t_0,t_1,\cdots)+\cdots\\
& Q_{n}(t)=\epsilon q^{(0)}(n_0,x_1,\cdots;t_0,t_1,\cdots)+
\epsilon^2 q^{(1)}(n_0,x_1,\cdots;t_0,t_1,\cdots)+\cdots
\end{align}
and the difference-differential operators must be understood as
\begin{equation}
\nabla_n^ \pm\to \nabla_{n_0}^ \pm +\epsilon\partial_{x_1}+\cdots\ ,\quad
\partial_t\to \partial_{t_0}+\epsilon\partial_{t_1}+\cdots\ .
\end{equation}
Hereabove the difference operators are defined as
\begin{equation}
\nabla_n^+\Psi_n=\Psi_{n+1}-\Psi_n\ ,\quad 
\nabla_n^-\Psi_n=\Psi_n-\Psi_{n-1},\end{equation}
such that the second order difference appearing in \eqref{dav-sco-psi} is
\begin{equation}
\nabla_n^2\Psi_n=\nabla_n^+\nabla_n^-\Psi_n=\Psi_{n+1}+\Psi_{n-1}-2\Psi_n.
\end{equation}

The system (\ref{dav-sco-psi}-\ref{dav-sco-q}) at first order gives 
the linear equations
\begin{align}
&{\cal L}_0\varphi^{(0)}=0\ ,\quad {\cal L}_0=i\partial_{t_0} +\nabla_{n_0}^2,
\nonumber\\
&L_0 q^{(0)}=0\ ,\quad L_0=\partial_{t_0}^2-V^2\nabla_{n_0}^2.
\label{linear}\end{align}
The next order $\epsilon^2$  eventually reads
\begin{align}
&\mathcal{L}_{0}\varphi^{(1)}=-i\partial_{t_1}\varphi^{(0)}
-(\nabla_{n_0}^+ +\nabla_{n_0}^-)\partial_{x_1}\varphi^{(0)}+
q^{(0)}\varphi^{(0)},\nonumber\\
&L_0q^{(1)}=2\partial_{t_0}\partial_{t_1}q^{(0)}
+V^2(\nabla_{n_0}^+ +\nabla_{n_0}^-)\partial_{x_1}q^{(0)}+
\nabla_{n_0}^2|\varphi^{(0)}|^2.
\label{nonlinear}\end{align}

The method then works as follows. Once selected an explicit solution of the
linear system \eqref{linear}, we express that the evolution \eqref{nonlinear}
of the first order correction $\{\varphi_n^{(1)},q_n^{(1)}\}$ must not produce
secular growth. This explicitely furnishes the evolution of the fundamental
$\{\varphi_n^{(0)},q_n^{(0)}\}$ in the slow variables $x_1$ and $t_1$.
The choice of the linear solution in the variables $n_0,t_0$ determines the
physical problem under study.

\subsection{Selection rules}

To describe a 3-wave interaction process involving incident and backscattered
optical phonon waves, we are led to select in the linear system \eqref{linear}
the solution
\begin{align}
&\varphi^{(0)}=A(x_1,t_1)e^ {i(k_1n_0-\omega_1t_0)}+ 
B(x_1,t_1)e^ {i(-k_2n_0-\omega_2t_0)}\ ,\\
&q^{(0)}=g(x_1,t_1)e^ {i(Kn_0-\Omega t_0)}+
\overline g(x_1,t_1) e^ {-i(Kn_0-\Omega t_0)}\ ,
\end{align}
with wave numbers $k_1>0$ and $k_2>0$ such as to ensure an optical phonon 
wave as a superposition of an incomming wave of amplitude $A$ and a reflected
wave of amplitude $B$.  The acoustic wave can propagate in both  directions,
thus $K$ can be of either signs. We have explicitely written the variables
$(x_1,t_1)$ in the slowly varying  amplitudes $A$, $B$, and $g$, but of course
they depend on all higher order  variables (but {\em not} on the first order
ones $n_0$ and $t_0$). 
The above expression is a solution of \eqref{linear} for the following 
dispersion relations 
\begin{align}\label{dispersion}
\omega_1=2(1-\cos k_1)\ ,\quad \omega_2=2(1-\cos k_2),\quad 
\Omega=2V|\sin\frac{K}{2}|.
\end{align}

Note that using the relation (\ref{Q-B}) and 
the multiscale derivative laws, we 
demonstrate that the first order enveloppe 
$\beta_0$ of the longitudinal displacement 
along the hydrogen-bond chain $\beta_n$ is related to the enveloppe 
$g(x_1,t_1)$ of $Q_n$ by
\begin{equation}
\beta_0(x_1,t_1)=-i\frac{J}{2\chi\sin(K)}g(x_1,t_1)
\end{equation} 
then the behaviour of the enveloppe $g(x_1,t_1)$ 
will automatically gives the behaviour of $\beta_0(x_1,t_1)$.

The resonant wave interaction results from a selection rule for the wave
parameters, obtained by examination of the nonlinear terms that occur
in \eqref{nonlinear}, namely
\begin{align*}
q^{(0)}\varphi^{(0)}= &Age^ {i(k_1+K)n_0}e^ {-i(\omega_1+\Omega)t_0}
+A\overline ge^ {i(k_1-K)n_0}e^ {-i(\omega_1-\Omega)t_0}\\
&+Bge^ {i(-k_2+K)n_0}e^ {-i(\omega_2+\Omega)t_0}
+B\overline ge^ {i(-k_2-K)n_0}e^ {-i(\omega_2-\Omega)t_0}.
\end{align*} 
Such terms will combine to either components of $\varphi^{(0)}$ and resonate
with corresponding factors in the left-hand-side of \eqref{nonlinear}. The
evolution of the envelopes will be then obtained by setting to zero the
coefficients of resonating terms.

\begin{figure}[ht]
\centerline{\epsfig{file=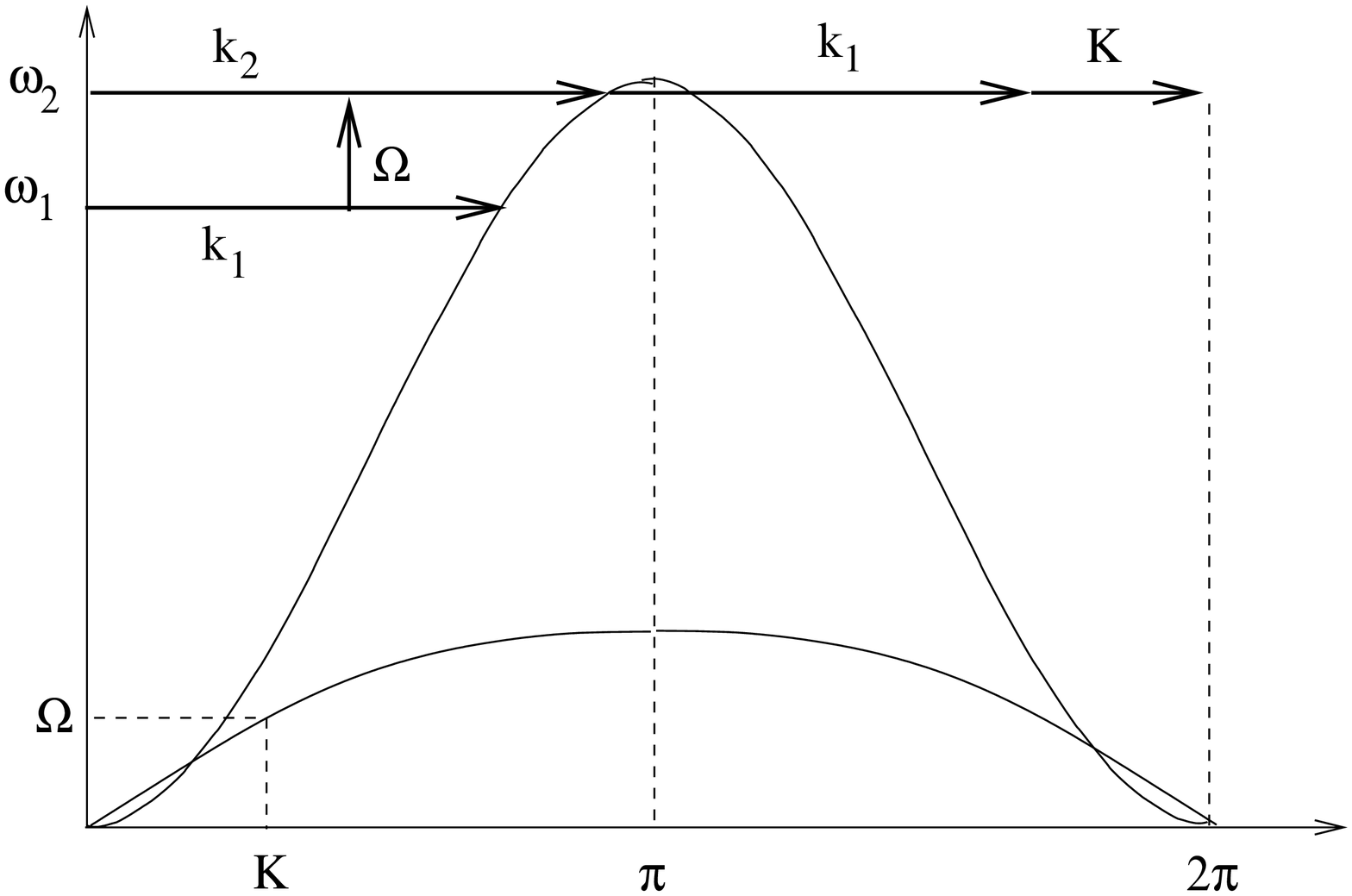,height=5cm,width=6cm}
\epsfig{file=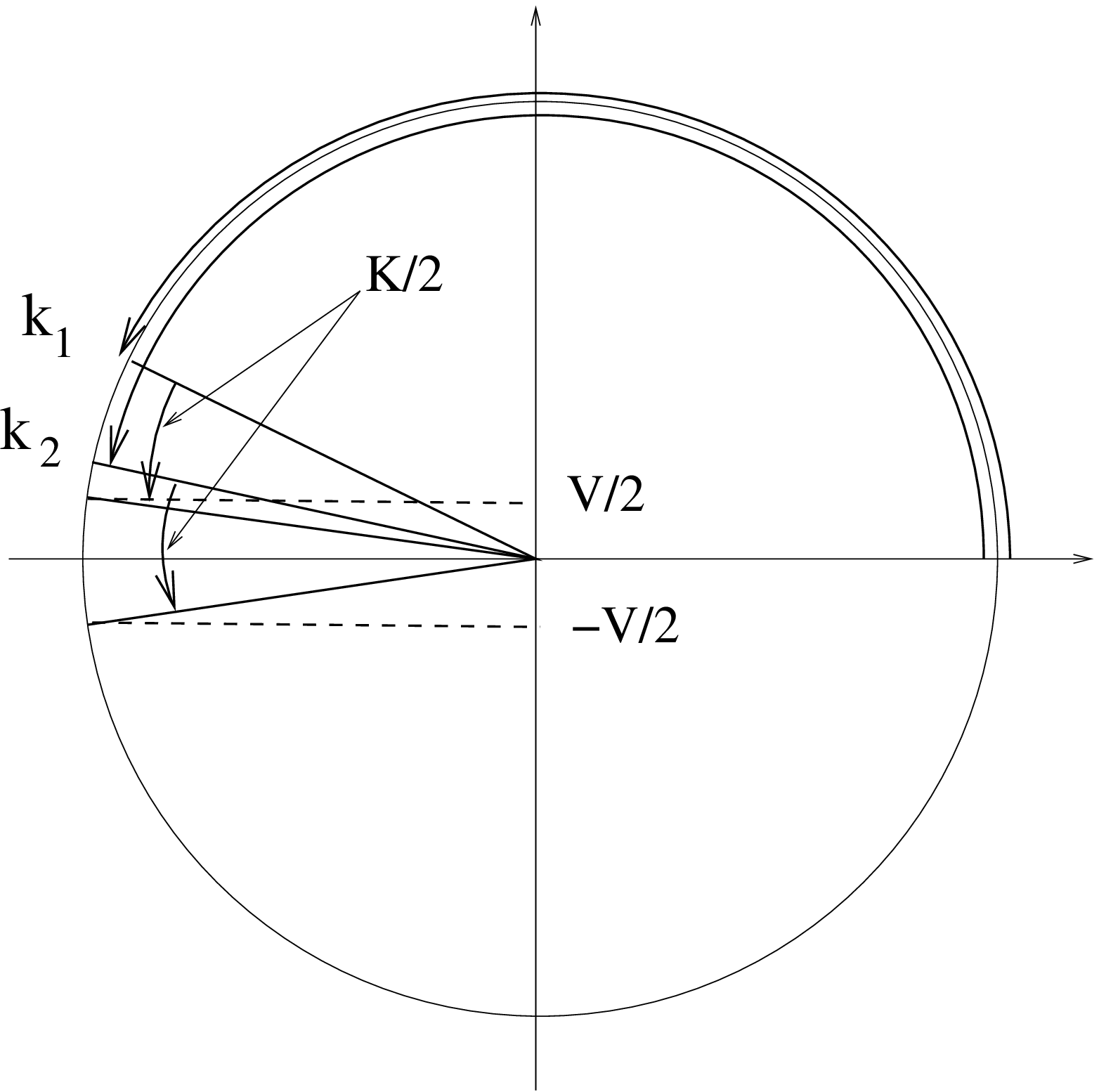,height=5cm,width=5cm}}
\caption {Representation of the selection rules \eqref{select1}  ($V=0.5$)
and graph of the solution \eqref{sol-select}.}
\label{fig:disp}\end{figure}

Since the physical context is the resonant interaction of two HF-waves (optical
phonon) with a LF acoustic wave, it implies a  small value of $K$ (near
the center of the Brillouin zone) and large values of $k_1$ and $k_2$, that is
to say close to, but less than, the value $\pi$ in order to achieve incident
and reflected HF-waves. We are then left with the following selection rules.
First when the nonlinear terms $Ag$ and $B\bar g$ combine respectively with
$B$ and $A$ we get
\begin{equation}
k1+k2=2\pi-K ,\quad K>0,\quad \omega_1-\omega_2=-\Omega,\label{select1}
\end{equation}
Another possibility is to combine instead $A\bar g$ and $Bg$ with
$B$ and $A$ to obtain
\begin{equation}
k1+k2=2\pi+K,\quad K<0 ,\quad\omega_1-\omega_2=\Omega.\label{select2}
\end{equation}
We shall study these two cases and demonstrate that only the first one gives
an instability that generates solitons, it is displayed on fig.~\ref{fig:disp}.

It is worth remarking that such a scattering process (involving $2\pi$) is 
allowed for by the presence of an exponential in the discrete variable ($n_0$),
in other words by the  discrete nature of the medium.

Inserting the dispersion relations \eqref{dispersion} in the equation
 \eqref{select1}, we
obtain after some algebra the following solution
\begin{equation}\label{sol-select}
2\sin (k_1+K/2)=V,\quad 2\sin (k_2+K/2)=-V.\end{equation}
These equations  determine completely the wave nubers $k_2$ and $K$ from the
data of $V$ (physics) and $k_1$ (incident wave), as soon as one assumes that
$k_1$ an $k_2$ are close to, and less than $\pi$.

Remark : a particular case is obtained in the limit $K\to 0$ for which $k_2\to
-k_1$ and $2\sin (k_1)=V$. This is the {\em two-wave} resonant interaction
studied in \cite{flora} for which a different multi-scale analysis has to be
employed (resonant process occurs at larger scales).

\subsection{Three-wave resonant interaction}

Corresponding to the first selection rule \eqref{select1}, the evolution
equations for the envelopes that ensure vanishing of the resonant terms in the
evolution  \eqref{nonlinear} read
\begin{align}
&[\partial_{t }+2\sin k_1\partial_{x }]A=-iB\overline g,\nonumber\\
&[\partial_{t }-2\sin k_2\partial_{x }]B=-iAg,\label{syst-1}\\
&[\partial_{t }-V^2\frac{\sin K}{\Omega}\partial_{x }]g=
i\frac {\Omega}{2V^2}B\overline A.\nonumber
\end{align} 
From now on we rename $x_1$ by $x$ and $t_1$ by $t$.
The same procedure applied to the second selection rule \eqref{select2} 
eventually furnishes
\begin{align}
&[\partial_{t }+2\sin k_1\partial_{x }]A=-iB g,\nonumber\\
&[\partial_{t }-2\sin k_2\partial_{x }]B=-iA\overline g\,\label{syst-2}\\
&[\partial_{t }-V^2\frac{\sin K}{\Omega}\partial_{x }]g=
i\frac {\Omega}{2V^2}A\overline B .\nonumber
\end{align} 
These are two standard 3-wave interaction nonlinear evolutions which are
integrable systems on the infinite line $x \in{\mathbb R}$ when Cauchy initial
data are prescribed  in a space of functions vanishing (together with all their
derivatives) at $x\to\pm\infty$ \cite{K1}.

Such an initial-boundary value problem is not the one we are interested in as
indeed we study  the scattering of an incident HF wave of envelope $A(x ,t )$
having a prescribed value for all time $t $ at the origin of the medium, i.e.
at $x =0$. Moreover, as $B(x ,t )$ stands for the envelope of the reflected
wave, its value will be prescribed to vanish at the output $x =L$ for all
time. Thus we cannot make use of the inverse scattering transform, unless first
reformulated for a boundary-value problem on the finite interval, which is
still an open problem.

Note that the system \eqref{syst-2}, by renaming $g$ as $\overline{g}$, maps to
the first one except for the sign of the inhomogeneous term of the last
equation. This change of sign is fundamental as it switches from instability to
stability, as described below.

\subsection{Stability properties}

The problem we consider  is thus the scattering of an incident HF wave of
envelope $A(x ,t )$, belonging to a carrier wave of frequency
$\omega_1$, that generates  backscattered wave of envelope $B(x ,t )$ and  LF
acoustic wave of envelope $g(x ,t )$ out of initial vacuum. Thus we perform a
linear stability analysis of both systems about the solution
\begin{equation}\label{CW-sol}
A(x ,t )=A_c\ ,\quad B=0\ ,\quad g=0.
\end{equation} 
with constant $A_c$ corresponding to a continuous wave (CW) irradiation.
Mathematically speaking this is an exact solution of both systems
\eqref{syst-1} and \eqref{syst-2} and thus only an instability could produce an
effective scattering. Let us seek now a solution as the perturbation
\begin{equation}
A=A_c+\epsilon\  a\ e^ {-i\nu t },\quad
B=\epsilon\  b \ e^ {-i\nu t },\quad
g=\epsilon\  q \ e^ {-i\nu t }.
\end{equation} 

The system \eqref{syst-1} at order $\epsilon$ then gives the linear system
\begin{equation}\label{I-I}
\left(\begin{array}{ccc} \nu&0&0\\ 0&\nu&A_c\\
0&-\bar{A_c}\frac{\Omega}{2V^2}&\nu \end{array}\right)
\left(\begin{array}{c}a\\ b\\ q \end{array}\right)=0
\end{equation} 
possessing the 3 eigenvalues $\nu=0$ and $\nu=\pm i(|A_c|/V)\sqrt{\Omega/2}$.
The system is thus unstable ($\Omega>0$), while the same analysis with system
\eqref{syst-2} and $g=\epsilon q e^ {i\nu t }$, furnishes the real
eigenvalues  $0$ and  $\pm(|A_c|/V)\sqrt{\Omega/2}$, and thus stability.

Consequently the linear stability analysis predicts that the selection rule
\eqref{select1} will produce an effective scattering for an incident CW wave. 
Our purpose now is to demonstrate by numerical simulations that this
instability is a {\em soliton generator} which induces an effective energy
absorption of the incident radiation.

\section{Numerical simulations}

\begin{figure}[ht]
\centerline{\epsfig{file=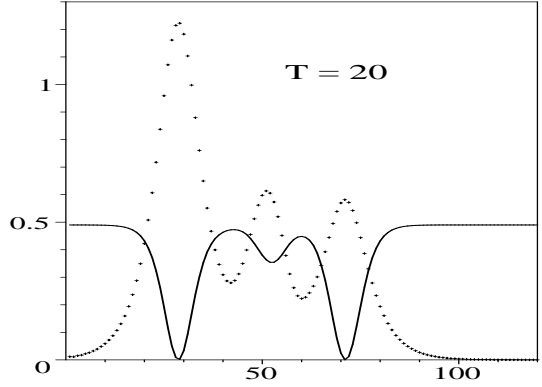,height=5cm,width=7cm}}
\caption{Energy density at time $t=20$ for optical phonon wave 
$|A(x,t)|^2$ (full line) and $|g(x,t)|^2$ (cross).}\label{fig:expe}
\end{figure}

In order to understand the effect of the instability of the solution 
\eqref{CW-sol} in system \eqref{syst-1}, and compare it to system
\eqref{syst-2}, we perform here numerical simulations of those sytems under the
following initial-boundary value data on the interval $x\in[0,L]$,
\begin{equation}
A(0,t)=A_c,\quad B(L,t)=0,\quad g(x,0)=0.\end{equation}
As an illustartion  the figure \ref{fig:expe} shows the energy density
profile $|A(x,t)|^2$ of the incident wave as a function of $x$ at time 
$t=20$ 
for $A_c=0.7$. The dashed line represents the acoustic phonon $|g(x,t)|^2$
that has gained the energy lost by the incident optical phonon.

\begin{figure}[ht]
\centerline{\epsfig{file=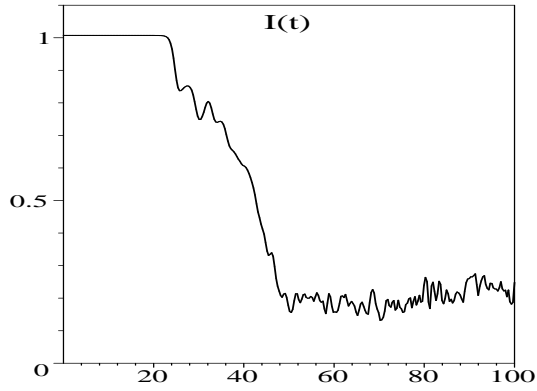,height=5cm,width=7cm}}
\caption{Energy  of the optical phonon wave 
$I(t)$  for $A_c=0.7$ and the values \eqref{param}}\label{fig:energy}
\end{figure}

This process corresponds to a brutal transfer of energy from the incident wave
(which has already settled in the medium) to the medium, as shown on
fig.~\ref{fig:energy}
The normalized energy carried by the waves of envelopes $A$ and $g$ are 
defined by
\begin{align}
&I(t)=\frac1{L|A_c^2|^2}{\int_{0}^Ldx\,|A(x,t)|^2},\\
&P(t)=\frac1{L|A_c^2|^2}{\int_{0}^Ldx\,|g(x,t)|^2}.
\end{align}
These are the quantities evaluated on a numerical simulation in 
fig.~\ref{fig:energy} for the following parameters
\begin{align}
&\omega_1=3.94 ,\quad k_1=2.89,\nonumber\\
&\omega_2=3.98,\quad  k_2=2.99,\nonumber\\
&V=0.1,\quad \Omega=0.04,\quad K=0.4.\label{param}\end{align}

Performing the same numerical simulations in the case of the system
\eqref{syst-2} have never produced any nonlinear energy transfer, confirming the
predictions of the linear stability analysis.

\section{Conclusion}

We have shown that a molecular chain allowing for wave coupling process, like
the Davydov  model, can present {\em nonlinear energy absorption} by resonant
interaction, coupling HF to LF waves, with selection rules \eqref{select1} that
are allowed only in discrete systems.  

The resulting governing equation, though being integrable (for a Cauchy initial
value problem on the infinite line) leads to an instability when driven by
boundary data on the finite interval. This instability is then the source of
soliton formation in the medium, and energy transfers from the incident
radiation to the medium excitation.

This process is clearly illustrated by numerical simulations and opens the way
to further studies in different contexts where the discreteness is known {\em a
priori} to play a fundamental role.

\end{document}